\documentclass[a4paper]{jpconf}

\usepackage{bm}

\begin{document}

\title{Solutions in the $(\frac{1}{2},0)\oplus(0,\frac{1}{2})$ Representation of the Lorentz Group}

\author{ Valeriy V. Dvoeglazov}

\address{UAF, Universidad Aut\'onoma de Zacatecas\\
Apartado Postal 636, Suc. 3, Zacatecas 98061 Zac., M\'exico}
%E-mail: valeri@fisica.uaz.edu.mx\\
%URL: http://fisica.uaz.edu.mx/\~{}valeri/\\
%}

\ead{valeri@fisica.uaz.edu.mx}

\date{\empty} 

%\maketitle

%\medskip

\begin{abstract}
I present explicit examples of generalizations in relativistic quantum mechanics.
First of all, I discuss the generalized spin-1/2 equations
for neutrinos. They have been obtained by means of the Gersten-Sakurai method
for derivations of arbitrary-spin relativistic equations. Possible
physical consequences are discussed.
Next, it is easy to check that both Dirac algebraic equation  $Det (\hat p - m) =0$ and $Det (\hat p + m) =0$  for  $u-$ and $v-$  4-spinors have solutions with $p_0= \pm E_p =\pm \sqrt{{\bf p}^2 +m^2}$. The same is true for higher-spin equations. Meanwhile, every book considers
 the equality $p_0=E_p$  for both $u-$  and $v-$ spinors of the $(1/2,0)\oplus (0,1/2))$ representation only, thus applying the Dirac-Feynman-Stueckelberg procedure for elimination of the negative-energy solutions. The recent work by Ziino (and, independently, 
the articles of several others) show that the Fock space can be doubled. We re-consider this possibility on the quantum field level for 
$S=1/2$ particles.
The third example is: we postulate the non-commutativity of 4-momenta, and we derive the mass splitting in the Dirac equation.
Some applications are discussed.
\end{abstract}

%\newpage

%\large{

\section{Generalized Neutrino Equations.}

A. Gersten~\cite{Ger} proposed a method for derivations
of massless equations of arbitrary-spin particles.
In fact, his method is related to the van der Waerden-Sakurai~\cite{Sak}
procedure for the derivation  of the massive Dirac equation. I commented on
the derivation of the Maxwell equations (the $S=1$ first-quantized
equations) in~\cite{dvo1}. Then I showed that one can obtain {\it
generalized} $S=1$ equations, which connect the antisymmetric tensor field
with additional scalar fields. The problem of physical significance of
additional scalar fields should be solved by experiment (see, however, 
the note on QED~\cite{Kruglov}).
In the present talk I apply the similar procedure to the spin-1/2 fields. As a result one obtains equations which
{\it generalize} the well-known Weyl equations. However, these equations
have been known for a long time~\cite{antec}. 
Raspini~\cite{Rasp1,Rasp5} analyzed them again in detail.
I add some comments on  physical
significance of the generalized spin-1/2 equations.

%\section{Derivation}

Let me use the equation (4) of the first Gersten paper~\cite{Ger}
for the two-component spinor field function:
\begin{equation}
(E^2 -c^2 \vec{\bf p}^{\,2}) I^{(2)}\psi =
\left [E I^{(2)} - c \vec {\bf p}\cdot \vec{\bm \sigma} \right ]
\left [E I^{(2)} + c \vec {\bf p}\cdot \vec{\bm \sigma} \right ]
\psi = 0\,.\label{G1}
\end{equation}
Actually, this equation is the massless limit of the equation which
was  presented in the Sakurai book~\cite{Sak}. In the latter case one
should substitute $m^2 c^4$ into the right-hand side of Eq.~(\ref{G1}):
\begin{equation}
\left [E I^{(2)} - c \vec {\bf p}\cdot \vec{\bm \sigma} \right ]
\left [E I^{(2)} + c \vec {\bf p}\cdot \vec{\bm \sigma} \right ]
\psi = m^2_2 c^4\psi\,.\label{G1A}
\end{equation}
However, instead of equation (3.25)
of~\cite{Sak} one can define the two-component `right' field function
\begin{equation}
\phi_R= {1\over m_1 c} (i\hbar {\partial \over \partial x_0}
-i\hbar {\bm \sigma}\cdot {\bm \nabla}) \psi,\quad\phi_L=\psi\,
\end{equation}
with the different mass parameter $m_1$.
In such a way we come to the system of the first-order differential equations
\begin{eqnarray}
&&(i\hbar {\partial \over \partial x_0}
+i\hbar {\bm \sigma}\cdot {\bm \nabla}) \phi_R
={m_2^2 c\over m_1}\phi_L\,,\\
&&(i\hbar {\partial \over \partial x_0}
-i\hbar {\bm \sigma}\cdot {\bm \nabla}) \phi_L
=m_1 c\phi_R\,.
\end{eqnarray}
It can be re-written
in the 4-component form:
\begin{eqnarray}
\label{gde}
&&\pmatrix{i\hbar (\partial/\partial x_0) &
i\hbar {\bm \sigma}\cdot {\bm \nabla}\cr
-i\hbar {\bm \sigma}\cdot {\bm \nabla}&
-i\hbar (\partial/\partial x_0)}
\pmatrix{\psi_A\cr\psi_B} =\\ 
&=&{c\over 2}
\pmatrix{(m_2^2/m_1
+m_1)&
(-m_2^2/m_1 +
m_1)\cr
(-m_2^2/m_1 +
m_1)& (m_2^2/m_1
+m_1)\cr}\pmatrix{\psi_A\cr\psi_B\cr}\nonumber
\end{eqnarray}
for the function $\Psi = column (\psi_A\quad \psi_B)=
column (\phi_R+\phi_L\quad \phi_R - \phi_L )$.
The equation (\ref{gde}) can be written in the covariant form:
\begin{equation}
\left [ i\gamma^\mu \partial_\mu - {m_2^2 c\over m_1 \hbar}
{(1-\gamma^5)\over 2} -{m_1 c \over \hbar} {(1+\gamma^5)\over 2}
\right ]\Psi = 0\,.\label{gde1}
\end{equation}
The standard representation of $\gamma^\mu$ matrices
has been used here.

If $m_1=m_2$ we can recover the standard Dirac equation.
As noted in~[5b] this procedure can be viewed as the simple change of
the representation of $\gamma^\mu$ matrices. However, this is valid only if the mass to not be equal to zero.
Otherwise, the entries in the transformation matrix become singular. The unitary  matrix does not exist.

Furthermore, one can  either repeat a similar procedure
(the modified Sakurai procedure) starting from the {\it massless}
equation (4) of~[1a] or put $m_2=0$ in eq.  (\ref{gde1}). The {\it massless
equation} is
\begin{equation}
\left [
i\gamma^\mu \partial_\mu - {m_1 c \over \hbar} {(1+\gamma^5)\over 2}
\right ]\Psi = 0\,.\label{gd1}
\end{equation}
It is necesary to stress that the term {\it `massless'} is used in the sense that $p_\mu p^\mu =0$.
The solutions of the equation (\ref{gd1}) are
\begin{eqnarray}
u_\sigma^{(1)} ({\bf p})= \pmatrix{[E+{\bm \sigma}\cdot {\bf p}]\phi_\sigma\cr
[E+m_1-{\bm \sigma}\cdot {\bf p}]\chi_\sigma\cr}\,,\quad v_\sigma^{(1)} ({\bf p} =\gamma^5 u_\sigma^{(1)} ({\bf p})\,.\label{s1}
\end{eqnarray}
in the spinorial representation of $\gamma$-matrices.
Then, we may have different physical consequences following from (\ref{gd1}) comparing with 
 those which follow from 
the Weyl equation.\footnote{Remember 
that the Weyl equation
is obtained as $m\rightarrow 0$ limit of the usual Dirac equation.} The
mathematical reason for such a possibility of  different massless
limits is that the corresponding change of representation of $\gamma^\mu$
matrices involves the mass parameters $m_1$ and $m_2$ themselves. 

It is interesting to note that we can also repeat this procedure
for the definition (or for even more general definitions):
\begin{equation}
\phi_L= {1\over m_3 c} (i\hbar {\partial \over \partial x_0}
+i\hbar {\bm \sigma}\cdot {\bm \nabla}) \psi,\quad\phi_R=\psi\,,
\end{equation}
with the additional arbitrary mass parameter $m_3$.
This is due to the fact that the parity properties of
the two-component spinor are undefined in the two-component equation.
The resulting equation is
\begin{equation}
\left [ i\gamma^\mu \partial_\mu - {m_4^2 c\over m_3 \hbar}
{(1+\gamma^5)\over 2} -{m_3 c \over \hbar} {(1-\gamma^5)\over 2}
\right ]\tilde\Psi = 0\,,
\end{equation}
which gives us  yet another equation in the massless limit (the physical mass $m_4
\rightarrow 0$):
\begin{equation} \left [ i\gamma^\mu \partial_\mu - {m_3
c \over \hbar} {(1-\gamma^5)\over 2} \right ]\tilde\Psi = 0\,, \label{gd2}
\end{equation}
differing in the sign at the $\gamma_5$ term. Solutions of the equation (\ref{gd2}) are
\begin{eqnarray}
u_\sigma^{(2)} ({\bf p})= \pmatrix{[E+m_3+{\bm \sigma}\cdot {\bf p}]\tilde\phi_\sigma\cr
[E-{\bm \sigma}\cdot {\bf p}]\tilde\chi_\sigma\cr}\,,\quad v_\sigma^{(2)} ({\bf p} =\gamma^5  u_\sigma^{(2)} ({\bf p})\,.\label{s2}
\end{eqnarray}
At this moment, neither (\ref{s1}) nor (\ref{s2}) are orthonormalized.

The above procedure can be generalized to {\it any} Lorentz group
representations, {\it i.~e.}, to any spins.
In some sense the equations (\ref{gd1}),(\ref{gd2}) are
analogous to the  $S=1$ equations~\cite[(4-7,10-13)]{dvo1}, which
also contain additional parameters.

%\section{Physical Interpretations and Conclusions}

The physical content of the generalized $S=1/2$
{\it massless} equations may be different from that of the Weyl equation.
The excellent discussion can be found
in~[5a,b]. First of all, the theory does {\it not} have chiral invariance. Those authors
call the additional parameters the measure of the degree of chirality.
Apart of this, Tokuoka introduced the concept of the gauge transformations
(not to confuse with phase transformations) for the 4-spinor fields. He
also found some strange properties of the  anti-commutation relations
(see \S 3 in~[5a] and cf.~\cite{Santhanam} and~\cite{Ahlu,Ahlustat}). And finally, the equations (\ref{gd1},\ref{gd2}) describe
{\it four} states, two of which correspond to the positive energy $p_0=\vert
{\bf p}\vert$, and two others correspond to the negative energy $p_0=-\vert
{\bf p}\vert$.

I just want to add the following to the discussion.
The operator of the {\it chiral-helicity} $\hat\eta = ({\bm\alpha}\cdot
\hat{\bf p})$ (in the spinorial representation) used in~[5b] does {\it not} commute, 
{\it e.g.}, with the Hamiltonian of the
equation~(\ref{gd1}):\footnote{Do not confuse with the Dirac Hamiltonian.}
\begin{equation} [{\cal H}, {\bm\alpha}\cdot \hat{\bf
p} ]_- = 2{m_1 c \over \hbar} {1-\gamma^5 \over 2} ({\bm \gamma}\cdot
\hat{\bf p})\, ,
\end{equation}
thus not having the common eigenstates.
For the eigenstates of the {\it chiral-helicity} the system of corresponding
equations can be read ($\eta=\uparrow, \downarrow$)
\begin{equation} i\gamma^\mu
\partial_\mu \Psi_\eta - {m_1 c\over \hbar}{1+\gamma^5 \over 2}\Psi_{-\eta}
=0 \, .  \end{equation} The conjugated eigenstates of the Hamiltonian
$\vert \Psi_\uparrow + \Psi_\downarrow >$ and
$\vert \Psi_\uparrow - \Psi_\downarrow >$
are connected, in fact, by $\gamma^5$ transformation
$\Psi \rightarrow \gamma^5 \Psi \sim ({\bm\alpha}\cdot \hat{\bf p})
\Psi$ (or $m_1\rightarrow -m_1$).  However, the $\gamma^5$ transformation
is related to the $PT$ ($t\rightarrow - t$ only) transformation,
which, in its turn, can be interpreted as the change of the sign of the energy, if one
accepts the Stueckelberg idea about antiparticles.  We associate $\vert
\Psi_\uparrow + \Psi_\downarrow >$ with the positive-energy eigenvalue of
the Hamiltonian and $\vert \Psi_\uparrow -
\Psi_\downarrow >$, with the negative-energy eigenvalue of the
Hamiltonian. Thus, the free chiral-helicity
massless eigenstates may oscillate one to another with the frequency
$\omega = E/\hbar$ (as the massive chiral-helicity eigenstates, see~[11a]
for details). Moreover, a special kind of interaction which is not
symmetric with respect to the chiral-helicity states (for instance, if
the left chiral-helicity eigenstates interact with the matter only) may induce
changes in the oscillation frequency, like in the Wolfenstein (MSW) formalism.

The question is: how can these frameworks be connected
with the Ryder method of derivation of relativistic wave equations~\cite{Ryder}, and
with the subsequent analysis of the problems of the choice of
normalization and that of the choice of  phase factors in the papers~\cite{Dvo2,Ahlu,Dvo3}?
However, the conclusion may be similar to that which was  achieved before: the
dynamical properties of the massless particles ({\it e.~g.}, neutrinos and photons)
may differ from those defined by the well-known Weyl and Maxwell equations~\cite{Dvo3,Dvonew}.

\section{Negative Energies in the Dirac Equation.}

The recent problems of superluminal neutrinos, negative mass-squared neutrinos, various schemes of oscillations including sterile neutrinos, e.~g.~\cite{n3}, etc require much attention.
Next, the problem of the lepton mass splitting ($e, \mu, \tau$) has long history~\cite{ms}. This suggests that something is missed in the foundations of relativistic quantum theories. Modifications seem to be necessary in the Dirac sea concept, and in the even more sophisticated 
Stueckelberg concept of the backward propagation in time. The Dirac sea concept is intrinsically related to the Pauli principle. However, the Pauli principle is intrinsically connected with the Fermi statistics and the anticommutation relations of fermions. Recently, the concept of {\it bi-orthonormality} has been proposed; the (anti) commutation relations and statistics are assumed to be different for {\it neutral} particles~\cite{Ahlu} (cf.~\cite{Gelfand}).

%One can speculate that they go off in the negative-energy sea, but due to some reasons (interaction?) they do not live 
%there (from our viewpoint), but return back
%(been expelled), thus showing us the new kind of oscillations on the Planck scale $\omega >> E/\hbar$, Ref.~\cite{Dvonew}.
%Perhaps, some of the neutrinos remain {\it sterile} even in our world.
%
%We propose the relevant modifications in the basics of the relativistic quantum theory below. However much work is still needed.

%\section{The General Framework and Connections with Previous Models}

We observe some interisting things related to the negative-energy concept. Usually, everybody uses the following definition of the field operator~\cite{Itzykson} in the pseudo-Euclidean metrics:
%At least, 3 methods of its derivation exist~\cite{Dirac,Sak,Ryder}:
%\begin{itemize}
%  \item the Dirac one (the Hamiltonian should be linear in $\partial/\partial x^i$, and be compatible with $E_p^2 -{\bf p}^2 c^2 =m^2 c^4$);
%  \item the Sakurai one (based on the equation $(E_p- {\bf \sigma} \cdot {\bf p}) (E_p+ {\bf \sigma} \cdot {\bf p}) \phi =m^2 \phi$);
%  \item the Ryder one (the relation between  2-spinors at rest is $\phi_R ({\bf 0}) = \pm \phi_L ({\bf 0})$, and boost application to them).
%\end{itemize}
\begin{equation}
\Psi (x) = \frac{1}{(2\pi)^3}\sum_h \int \frac{d^3 {\bf p}}{2E_p} [ u_h ({\bf p}) a_h ({\bf p}) e^{-ip\cdot x}
+ v_h ({\bf p}) b_h^\dagger ({\bf p})] e^{+ip\cdot x}]\,,
\end{equation}
as given {\it ab initio}.
After actions of the Dirac operator on \linebreak  $\exp (\mp ip\cdot x)$ the 4-spinors ( $u-$ and $v-$ ) 
satisfy the momentum-space equations: $(\hat p - m) u_h (p) =0$ and $(\hat p + m) v_h (p) =0$, respectively; $h$ is 
the polarization index. However, it is easy to prove from the characteristic equations
$Det (\hat p \mp m) =(p_0^2 -{\bf p}^2 -m^2)^2= 0$ that the solutions should satisfy the energy-momentum relation $p_0= \pm E_p =\pm \sqrt{{\bf p}^2 +m^2}$ in both cases.

Let me recall the general scheme of construction of the field operator, which  was presented in~\cite{Bogoliubov}. In the case of
the $(1/2,0)\oplus (0,1/2)$ representation we have:
\begin{eqnarray}
&&\Psi (x) = {1\over (2\pi)^3} \int d^4 p \,\delta (p^2 -m^2) e^{-ip\cdot x}
\Psi (p) =\nonumber\\
&=& {1\over (2\pi)^3} \sum_{h}^{}\int d^4 p \, \delta (p_0^2 -E_p^2) e^{-ip\cdot x}
u_h (p_0, {\bf p}) a_h (p_0, {\bf p}) =\label{fo}\\
&=&{1\over (2\pi)^3} \int {d^4 p \over 2E_p} [\delta (p_0 -E_p) +\delta (p_0 +E_p) ] \nonumber\\
&&[\theta (p_0) +\theta (-p_0) ] e^{-ip\cdot x}
\sum_{h}^{} u_h (p) a_h (p) =\nonumber\\
&=& {1\over (2\pi)^3} \sum_h^{} \int {d^4 p \over 2E_p} [\delta (p_0 -E_p) +\delta (p_0 +E_p) ] \nonumber\\\
&&\left
[\theta (p_0) u_h (p) a_h (p) e^{-ip\cdot x}  + 
\theta (p_0) u_h (-p) a_h (-p) e^{+ip\cdot x} \right ]=\nonumber\\
&=& {1\over (2\pi)^3} \sum_h^{} \int {d^3 {\bf p} \over 2E_p} \theta(p_0)  
\left [ u_h (p) a_h (p)\vert_{p_0=E_p} e^{-i(E_p t-{\bf p}\cdot {\bf x})}  +\right.\nonumber\\ 
&+& \left. u_h (-p) a_h (-p)\vert_{p_0=E_p} e^{+i (E_p t- {\bf p}\cdot {\bf x})} 
\right ]\nonumber
\end{eqnarray}
During the calculations above we had to represent $1=\theta (p_0) +\theta (-p_0)$
in order to get positive- and negative-frequency parts.\footnote{See Ref.~\cite{DvoeglazovJPCS}
for some discussion.} Moreover, during these calculations we did not yet assume, which equation this
field operator  (namely, the $u-$ spinor) does satisfy, with negative- or positive- mass (energy).
In general we should transform $u_h (-p)$ to the $v (p)$. The procedure is the following one~\cite{DvoeglazovHJ}.
In the Dirac case we should assume the following relation in the field operator:
\begin{equation}
\sum_{h=\pm 1/2}^{} v_h (p) b_h^\dagger (p) = \sum_{h=\pm 1/2}^{} u_h (-p) a_h (-p)\,.\label{dcop}
\end{equation}
We know that~\cite{Ryder}\footnote{${(\mu)}$ and ${(\lambda)}$ are the polazrization indices here. According to 
the referee advice I use parenthesis here to stress this.}
\begin{eqnarray}
\bar u_{(\mu)} (p) u_{(\lambda)} (p) &=& +m \delta_{\mu\lambda}\,,\\
\bar u_{(\mu)} (p) u_{(\lambda)} (-p) &=& 0\,,\\
\bar v_{(\mu)} (p) v_{(\lambda)} (p) &=& -m \delta_{\mu\lambda}\,,\\
\bar v_{(\mu)} (p) u_{(\lambda)} (p) &=& 0\,,
\end{eqnarray}
but we need $\Lambda_{(\mu)(\lambda)} (p) = \bar v_{(\mu)} (p) u_{(\lambda)} (-p)$.
By direct calculations,  we find
\begin{equation}
-mb_{(\mu)}^\dagger (p) = \sum_{\lambda}^{} \Lambda_{(\mu)(\lambda)} (p) a_{(\lambda)} (-p)\,.
\end{equation}
Hence, $\Lambda_{(\mu)(\lambda)} = -im ({\bm \sigma}\cdot {\bf n})_{(\mu)(\lambda)}$, ${\bf n} = {\bf p}/\vert{\bf p}\vert$, 
and 
\begin{equation}
b_{(\mu)}^\dagger (p) = +i\sum_\lambda ({\bm\sigma}\cdot {\bf n})_{(\mu)(\lambda)} a_{(\lambda)} (-p)\,.
\end{equation}
Multiplying (\ref{dcop}) by $\bar u_{(\mu)} (-p)$ we obtain
\begin{equation}
a_{(\mu)} (-p) = -i \sum_{\lambda} ({\bm \sigma} \cdot {\bf n})_{(\mu)(\lambda)} b_{(\lambda)}^\dagger (p)\,.
\end{equation}
The equations are self-consistent.\footnote{In the $(1,0)\oplus (0,1)$ representation 
the similar procedure leads to somewhat different situation:
\begin{equation}
a_{(\mu)} (p) = [1-2({\bf S}\cdot {\bf n})^2]_{(\mu)(\lambda)} a_{(\lambda)} (-p)\,. 
\end{equation}
This signifies that in order to construct the Sankaranarayanan-Good field operator (which was used recently), 
it satisfies 
$[\gamma_{\mu\nu} \partial_\mu \partial_\nu - {(i\partial/\partial t)\over E} 
m^2 ] \Psi (x) =0$, we need additional postulates. For instance, one can try to construct 
the left- and the right-hand side of the field operator separately each other~\cite{DvoeglazovJPCS}.}

However, other ways of thinking are possible. First of all to mention, we have, in fact,
$u_h ( E_p, {\bf p} )$ and $u_h (-E_p, {\bf p})$, and $v_h ( E_p, {\bf p} )$ and $v_h (-E_p, {\bf p})$ originally, 
which satisfy the equations:\footnote{Remember that, as before, we can always make the substitution 
${\bf p}\rightarrow -{\bf p}$ in any of the integrands of (\ref{fo}).}
\begin{equation}
\left [ E_p (\pm \gamma^0) - {\bm \gamma}\cdot {\bf p} - m \right ] u_h (\pm E_p, {\bf p})=0\,.
\end{equation}
Due to the properties $U^\dagger \gamma^0 U=-\gamma^0$, $U^\dagger \gamma^i U=+\gamma^i$
with the unitary matrix $U=\pmatrix{0&-1\cr 1&0\cr}= \gamma^0\gamma^5$ in 
the Weyl basis,\footnote{The properties of the $U$-matrix are opposite to those of
$P^\dagger \gamma^0 P=+\gamma^0$, $P^\dagger \gamma^i P=-\gamma^i$
with the usual $P=\gamma^0$, thus giving $\left [ -E_p \gamma^0 + {\bf \gamma}\cdot {\bf p} - m \right ] 
P u_h (- E_p, {\bf p}) = -\left [\hat p +m \right ] \tilde v_{?} (E_p, {\bf p}) = 0$. While, the relations of the spinors $v_h (E_p, {\bf p})=\gamma_5  u_h (E_p, {\bf p})$ are well-known, it seems that
the relations of the $v-$ spinors of the positive energy to $u-$ spinors of the negative energy
are frequently forgotten, $\tilde v_{?} (E_p, {\bf p}) = \gamma^0 u_h (- E_p, {\bf p})$. Bogoliubov 
and Shirkov~\cite[p.55-56]{Bogoliubov} used  to construct the complete set
of solutions of the relativistic equations, fixing the sign of $p_0=+E_p$.} 
we have
\begin{equation}
\left [ E_p \gamma^0 - {\bm \gamma}\cdot {\bf p} - m \right ] U^\dagger u_h (- E_p, {\bf p})=0\,.\label{nede}
\end{equation}
Thus, unless the unitary transformations do not change
the physical content, we have that the negative-energy spinors $\gamma^5 \gamma^0 u^-$ (see (\ref{nede})) satisfy the accustomed ``positive-energy" Dirac equation. We should then expect the same physical content. Their explicit forms $\gamma^5 \gamma^0 u^-$ are different from the textbook ``positive-energy" Dirac spinors, while, of course, they should be  superpositions of the latter.
They are the following ones:
\begin{eqnarray}
\tilde u_1 (p) = \frac{N}{\sqrt{2m (-E_p +m)}} \pmatrix{-p^+ + m\cr -p_r\cr
p^- -m \cr - p_r\cr}\,,\\
\tilde u_2 (p) =\frac{N}{\sqrt{2m (-E_p +m)}}\pmatrix{-p_l \cr -p^- + m\cr
-p_l \cr p^+ -m\cr}\,.
\end{eqnarray}
$E_p=\sqrt{{\bf p}^2 +m^2}>0$, $p_0=\pm E_p$, $p^\pm = E_p\pm p_z$, $p_{r,l}= p_x\pm ip_y$.
Their normalization is to $(-2N^2)$.
What about the $\tilde v (p)=\gamma^0 u^-$ transformed with the $\gamma_0$ matrix?
They are not equal to $ v_h (p) =\gamma^5 u_h (p)$.
Obviously, they also do not have well-known forms  of the usual $v-$ spinors in the Weyl basis, 
differing by phase factors and in the signs at the mass terms. Their transformation properties are different:
\begin{eqnarray}
\tilde v_\tau ({\bf p}) = -i \frac{({\bm \sigma}^\ast\cdot {\bf p})_{\tau\sigma}}{p} v_\sigma ({\bf p})\,.
\end{eqnarray}

Next, one can prove that the matrix
\begin{equation}
P= e^{i\theta}\gamma^0 = e^{i\theta}\pmatrix{0& 1_{2\times 2}\cr 1_{2\times 2} & 0\cr}
\label{par}
\end{equation}
can be used in the parity operator as well as
in the original Weyl basis. The parity-transformed function
$\Psi^\prime (t, -{\bf x})=P\Psi (t,{\bf x})$ must satisfy
\begin{equation}
[i\gamma^\mu \partial_\mu^{\,\prime} -m ] \Psi^\prime (t,-{\bf x}) =0 \,,
\end{equation}
with $\partial_\mu^{\,\prime} = (\partial/\partial t, -{\bf \nabla}_i)$.
This is possible when $P^{-1}\gamma^0 P = \gamma^0$ and
$P^{-1} \gamma^i P = -\gamma^i$. The matrix (\ref{par})
satisfies these requirements, as in the textbook case.
However, if we would take the phase factor to be zero
we obtain that while spinors $u_h (p)$ have the eigenvalues $+1$ of the parity, but ($R= ({\bf x} \rightarrow -{\bf x}, {\bf p} 
\rightarrow -{\bf p}$))
\begin{equation}
P R\tilde u (p) = P R\gamma^5 \gamma^0  u (-E_p, {\bf p})= -\tilde u (p)\,,\quad 
\end{equation}
%\begin{equation}
%P R \tilde{\tilde u} (p) = P R \gamma^5 \gamma^0  u (-E_p, {\bf p})= -\tilde{\tilde u} (p)\,.
%\end{equation}
Perhaps, one should choose the phase factor $\theta=\pi$. Thus, we again confirm
that only the relative (particle-antiparticle) intrinsic parity has physical significance.

Similar formulations have been  presented in Refs.~\cite{Markov}, 
and~\cite{BarutZiino}. Namely, the reflection properties are different for some solutions 
of relativistic equations therein. Two opposite signs at the mass terms have been taken into account.
The group-theoretical basis for such doubling has been given
in the papers by Gelfand, Tsetlin and Sokolik~\cite{Gelfand}, who first presented 
the theory of 5-dimensional spinors (or, the one in the 2-dimensional projective representation of the inversion group) in 1956 
(later called as ``the Bargmann-Wightman-Wigner-type quantum field theory" in 1993). The corresponding connection 
with the time reversion has been clarified therein. It was one of the first attempts to explain the $K$-meson decays.
M. Markov proposed  {\it two} Dirac equations with  the opposite signs at the mass term~\cite{Markov} to be taken into account:
\begin{eqnarray}
\left [ i\gamma^\mu \partial_\mu - m \right ]\Psi_1 (x) &=& 0\,,\\
\left [ i\gamma^\mu \partial_\mu + m \right ]\Psi_2 (x) &=& 0\,.
\end{eqnarray}
In fact, he studied all properties of this relativistic quantum model (while the quantum
field theory has not yet been completed in 1937). Next, he added and  subtracted these equations. What did he obtain?
\begin{eqnarray}
i\gamma^\mu \partial_\mu \varphi (x) - m \chi (x) &=& 0\,,\\
i\gamma^\mu \partial_\mu \chi (x) - m \varphi (x) &=& 0\,.
\end{eqnarray}
Thus, the corresponding $\varphi$ and $\chi$ solutions can be presented as some superpositions of the Dirac 4-spinors $u-$ and $v-$.
These equations, of course, can be identified with the equations for the Majorana-like $\lambda -$ and $\rho -$, which we presented 
in Ref.~[11a].\footnote{Of course, the signs at the mass terms
depend on, how do we associate the positive- or negative- frequency solutions with $\lambda$ and $\rho$.}
\begin{eqnarray}
i \gamma^\mu \partial_\mu \lambda^S (x) - m \rho^A (x) &=& 0 \,,
\label{11}\\
i \gamma^\mu \partial_\mu \rho^A (x) - m \lambda^S (x) &=& 0 \,,
\label{12}\\
i \gamma^\mu \partial_\mu \lambda^A (x) + m \rho^S (x) &=& 0\,,
\label{13}\\
i \gamma^\mu \partial_\mu \rho^S (x) + m \lambda^A (x) &=& 0\,.
\label{14}
\end{eqnarray}
Neither of them can be regarded as the Dirac equation.
However, they can be written in the 8-component form as follows:
\begin{eqnarray}
\left [i \Gamma^\mu \partial_\mu - m\right ] \Psi_{_{(+)}} (x) &=& 0\,,
\label{psi1}\\
\left [i \Gamma^\mu \partial_\mu + m\right ] \Psi_{_{(-)}} (x) &=& 0\,,
\label{psi2}
\end{eqnarray}
with
\begin{eqnarray}
&&\hspace{-20mm}\Psi_{(+)} (x) = \pmatrix{\rho^A (x)\cr
\lambda^S (x)\cr},
\Psi_{(-)} (x) = \pmatrix{\rho^S (x)\cr
\lambda^A (x)\cr}, \,\Gamma^\mu =\pmatrix{0 & \gamma^\mu\cr
\gamma^\mu & 0\cr}.
\end{eqnarray}
It is possible to find the corresponding Lagrangian, projection operators, and the Feynman-Dyson-Stueckelberg propagator.
For example,
\begin{eqnarray}
{\cal L} &=&\frac{i}{2} \left . [\overline\Psi_{(+)} \Gamma^{\mu}\partial_\mu \Psi_{(+)} - (\partial_\mu \overline\Psi_{(+)}) \Gamma^{\mu}
\Psi_{(+)} + \right .\\
&+&\left . \overline\Psi_{(-)} \Gamma^{\mu}\partial_\mu \Psi_{(-)} - (\partial_\mu \overline\Psi_{(-)}) \Gamma^{\mu}
\Psi_{(-)} \right ] - m [\overline\Psi_{(+)}\Psi_{(+)} - \overline\Psi_{(-)}\Psi_{(-)}]\,.\nonumber
\end{eqnarray}
The projection operator $P_+$ can be easily found, as usual,
\begin{equation}
P_+ = \frac{\Gamma_\mu p^\mu +m}{2m}\,.
\end{equation}
However, due to the fact that $P_-$ satisfies the Dirac equation with the opposite sign, we cannot have $P_+ + P_- =1$.
This is not surprising  because the corresponding states $\Psi_\pm$ do not form the complete system of the 8-dimensional space.
One should consider the states $\Gamma_5 \Psi_\pm ({\bf p})$ too.
See also~\cite{DVO-FD} for the methods of obtaining the propagators in the non-trivial cases.

In the previous papers I explained: the connection with the Dirac spinors has 
been found~\cite{Dvo2,Kirchbach} through the unitary matrix.
For instance,
\begin{eqnarray}
\pmatrix{\lambda^S_\uparrow ({\bf p}) \cr \lambda^S_\downarrow ({\bf p}) \cr
\lambda^A_\uparrow ({\bf p}) \cr \lambda^A_\downarrow ({\bf p})\cr} = {1\over
2} \pmatrix{1 & i & -1 & i\cr -i & 1 & -i & -1\cr 1 & -i & -1 & -i\cr i&
1& i& -1\cr} \pmatrix{u_{+1/2} ({\bf p}) \cr u_{-1/2} ({\bf p}) \cr
v_{+1/2} ({\bf p}) \cr v_{-1/2} ({\bf p})\cr},\label{connect}
\end{eqnarray}
provided that the 4-spinors have the same physical dimension.\footnote{The reasons of the change of the fermion 
mass dimension are unclear in the recent works on {\it elko}.}
Thus, this represents
itself the rotation of the spin-parity basis. However, it is usually assumed that the $\lambda-$ and $\rho-$ spinors describe the neutral particles,
meanwhile, the $u-$ and $v-$ spinors describe the charged particles. Kirchbach~\cite{Kirchbach} found the amplitudes for 
neutrinoless double beta decay ($00\nu\beta$) in this scheme. It is obvious from (\ref{connect}) that there are some additional terms comparing with the standard calculations of those amplitudes. 
One can also re-write the above equations into the two-component forms. Thus, one obtains the Feynman-Gell-Mann~\cite{FG} 
equations.

Barut and Ziino~\cite{BarutZiino} proposed yet another model. They considered
$\gamma^5$ operator as the operator of the charge conjugation. In their case the self/anti-self charge conjugate states
are, at the same time, the eigenstates of the chirality. Thus, the charge-conjugated
Dirac equation has a different sign compared with the ordinary formulation:
\begin{equation}
[i\gamma^\mu \partial_\mu + m] \Psi_{BZ}^c =0\,,
\end{equation}
and the so-defined charge conjugation applies to the whole system,  fermion + electro\-magnetic field, $e\rightarrow -e$
in the covariant derivative. The superpositions of the $\Psi_{BZ}$ and $\Psi_{BZ}^c$ also give us 
the ``doubled Dirac equation", as the equations for $\lambda-$ and $\rho-$ spinors. 
The concept of the doubling of the Fock space has been
developed in the Ziino works (cf.~\cite{Gelfand,DvoeglazovBW}) in the framework of the quantum field theory~\cite{Ziinonew}. 
Next, it is interesting to note that we have for the Majorana-like field operators ($a_\eta ({\bf p}) = b_\eta ({\bf p})$):
\begin{eqnarray}
\lefteqn{\left [ \nu^{^{ML}} (x^\mu) + {\cal C} \nu^{^{ML\,\dagger}} (x^\mu) \right ]/2 = 
\int {d^3 {\bf p} \over (2\pi)^3 } {1\over 2E_p} }\\
&&\sum_\eta \left [ \pmatrix{i\Theta \phi_{_L}^{\ast \, \eta} (p^\mu) \cr 0\cr} a_\eta
(p^\mu)  e^{-ip\cdot x} 
+\pmatrix{0\cr
\phi_L^\eta (p^\mu)\cr } a_\eta^\dagger (p^\mu) e^{ip\cdot x} \right ]\,,\nonumber
\\
\lefteqn{\left [\nu^{^{ML}} (x^\mu) - {\cal C} \nu^{^{ML\,\dagger}} (x^\mu) \right
]/2 = \int {d^3 {\bf p} \over (2\pi)^3 } {1\over 2E_p}} \\
&&\sum_\eta \left
[\pmatrix{0\cr \phi_{_L}^\eta (p^\mu) \cr } a_\eta (p^\mu)  e^{-ip\cdot x}
+\pmatrix{-i\Theta \phi_{_L}^{\ast\, \eta} (p^\mu)\cr 0
\cr } a_\eta^\dagger (p^\mu) e^{ip\cdot x} \right ]\, . \nonumber
\end{eqnarray}
This naturally leads to the Ziino-Barut scheme of massive chiral
fields, Ref.~\cite{BarutZiino}. See, however, the recent paper~\cite{DVO2018} which deals with the problems 
of the Majorana field operator.

Finally, I would like to mention that, in general,  in the Weyl basis the $\gamma-$ matrices are  {\it not} Hermitian, 
$\gamma^{\mu^\dagger}  =\gamma^0 \gamma^\mu \gamma^0$. So, $\gamma^{i^\dagger}= -\gamma^i$, $i=1,2,3$, the pseudo-Hermitian matrix.
The energy-momentum operator $i\partial_\mu$ is obviously Hermitian. So, the question, if the eigenvalues of 
the Dirac operator $i\gamma^\mu \partial_\mu$
(the mass, in fact) would be always real? The question of the complete system of the eigenvectors of the 
{\it non-}Hermitian operator deserve careful consideration~\cite{Ilyin}.   

%\section{Conclusions}       % Enter section title between curly braces

The main points of this Section are: there are ``negative-energy solutions" in that is previously  considered as 
``positive-energy solutions" of relativistic wave equations, and vice versa. Their explicit forms have been presented
in the case of spin-1/2. 
Next, relations to previous works have been found. For instance,
the doubling of the Fock space and the corresponding solutions of the Dirac equation
obtained  additional mathematical bases. Similar conclusion can be deduced for
the higher-spin equations. 

\section{Non-commutativity in the Dirac equation.}

The non-commutativity~\cite{snyder,amelino} manifests interesting peculiarities in
the Dirac case. We analized Sakurai-van der Waerden method of derivations of the Dirac
(and higher-spins too) equation~\cite{Dvoh}. We can start from
\begin{equation}
(E I^{(2)}-{\bm \sigma}\cdot {\bf p}) (E I^{(2)}+ {\bm\sigma}\cdot
{\bf p} ) \Psi_{(2)} = m^2 \Psi_{(2)} \,,
\end{equation}
or (in the 4-component case)
\begin{equation}
(E I^{(4)}+{\bm \alpha}\cdot {\bf p} +m\beta) (E I^{(4)}-{\bm\alpha}\cdot
{\bf p} -m\beta ) \Psi_{(4)} =0\,.\label{f4}
\end{equation}
Obviously, the inverse operators of the Dirac operators of the positive- and negative- masses exist  in the non-commutative case too.
As in the original Dirac work, we have
\begin{equation}
\beta^2 = 1\,,\quad
\alpha^i \beta +\beta \alpha^i =0\,,\quad
\alpha^i \alpha^j +\alpha^j \alpha^i =2\delta^{ij} \,.
\end{equation}
For instance, their explicite forms can be chosen 
\begin{eqnarray}
\alpha^i =\pmatrix{\sigma^i& 0\cr
0&-\sigma^i\cr}\,,\quad
\beta = \pmatrix{0&1_{2\times 2}\cr
1_{2\times 2} &0\cr}\,,
\end{eqnarray}
where, again, $\sigma^i$ are the ordinary Pauli $2\times 2$ matrices.

We postulate the non-commutativity relations for the components of 4-momenta:
\begin{equation}
[E, {\bf p}^i]_- = \Theta^{0i} = \theta^i\,,
\end{equation}
as usual. Therefore the equation (\ref{f4}) will {\it not} lead
to the well-known equation $E^2 -{\bf p}^2 = m^2$. Instead, we have
\begin{eqnarray}
&&\left \{ E^2 - E ({\bm \alpha} \cdot {\bf p})
+({\bm \alpha} \cdot {\bf p}) E - {\bf p}^2 - m^2 - i (I_{(2)}\otimes {\bm\sigma})
[{\bf p}\times {\bf p}] \right \} \Psi_{(4)} \nonumber\\
&&= 0
\end{eqnarray}
For the sake of simplicity, we may assume the last term to be zero. Thus, we come to
\begin{equation}
\left \{ E^2 - {\bf p}^2 - m^2 -  ({\bm \alpha}\cdot {\bm \theta})
\right \} \Psi_{(4)} = 0\,.
\end{equation} 
However, let us apply a unitary transformation. It is known, Refs.~\cite{Berg,Dvo2},
that one can transform\footnote{Some relations for the components ${\bf a}$ should be assumed. Moreover, in our case ${\bm \theta}$ 
should not depend on $E$ and ${\bf p}$. Otherwise, we must take the non-commutativity $[E, {\bf p}^i]_- \neq 0$ into account again.}
\begin{equation}
U_1 ({\bm \sigma}\cdot {\bf a}) U_1^{-1} = \sigma_3 \vert {\bf a} \vert\,.\label{s3}
\end{equation}
For ${\bm \alpha}$ matrices we re-write (\ref{s3}) to
\begin{eqnarray}
{\cal U}_1 ({\bm \alpha}\cdot {\bm \theta}) {\cal U}_1^{-1} = \vert {\bm \theta} \vert
\pmatrix{1&0&0&0\cr
0&-1&0&0\cr
0&0&-1&0\cr
0&0&0&1\cr} = \alpha_3 \vert {\bm\theta}\vert\,.
\end{eqnarray}
The explicit form of the $U_1$ matrix is ($a_{r,l}= a_1\pm ia_2$):
\begin{eqnarray}
U_1 &=&\frac{1}{\sqrt{2a (a+a_3)}} \pmatrix{a+a_3&a_l\cr
-a_r&a+a_3\cr}  = \frac{1}{\sqrt{2a (a+a_3)}} \nonumber\\
&\times&[ a+a_3 + i\sigma_2 a_1 - i\sigma_1 a_2]\,,\nonumber\\ 
{\cal U}_1 &=&\pmatrix{U_1 &0\cr
0& U_1 \cr}\,.
\end{eqnarray}
Let me apply the second unitary transformation:
\begin{eqnarray}
{\cal U}_2 \alpha_3 {\cal U}_2^\dagger &=&
\pmatrix{1&0&0&0\cr
0&0&0&1\cr
0&0&1&0\cr
0&1&0&0\cr} \alpha_3 \pmatrix{1&0&0&0\cr
0&0&0&1\cr
0&0&1&0\cr
0&1&0&0\cr} = \pmatrix{1&0&0&0\cr
0&1&0&0\cr
0&0&-1&0\cr
0&0&0&-1\cr}.\nonumber\\
&&
\end{eqnarray}
The final equation is
\begin{equation}
[E^2 -{\bf p}^2 -m^2 -\gamma^5_{chiral} \vert {\bm \theta}\vert ] \Psi^\prime_{(4)} = 0\,.\label{split}
\end{equation}
In the physical sense this implies the mass splitting for a Dirac particle over the non-commutative space, 
$m_{1,2} =\pm \sqrt{m^2 \pm \theta}$. 
This procedure may be attractive for explanation of the mass creation and the mass splitting for fermions.
One can also use the non-commutativity
\begin{equation}
[ {\bf p}^i, {\bf p}^j ]= \Xi^{ij}= \epsilon^{ijk}\xi^k
\end{equation}
with the corresponding substitutions: $\theta^i =0$, ${\cal U}_1 (\theta) \rightarrow ´{\cal U}_1 (\xi)$
and 
\begin{equation}
{\cal U'}_2 =
\pmatrix{1&0&0&0\cr
0&0&-1&0\cr
0&1&0&0\cr
0&0&0&1\cr}\,.
\end{equation}
In such a way we obtain the same splitting as in (\ref{split}), $\vert {\bm \theta}\vert \rightarrow \vert {\bm \xi}\vert$.

%}

\medskip

{\bf Acknowledgments.}
I greatly appreciate old discussions with Prof. A. Raspini
and useful information from Prof. A.~F.~Pash\-kov. I acknowledge discussions 
with participants of several recent Conferences.
This work has been partly supported by
the ESDEPED UAZ, M\'exico.

\end{document}